# The early evolution of solar flaring plasma loops

Baolin Tan[1,2]

[1] Key Laboratory of Solar Activity, National Astronomical Observatories of Chinese Academy of Sciences, Beijing 100012, China; bltan@nao.cas.cn

[2] School of Astronomy and Space Science, University of Chinese Academy of Sciences, Beijing 100049, China

**Abstract**: Plasma loops are the elementary structures of solar flaring active regions which dominate the whole processes of flaring eruptions. The standard flare models are well explained the evolution and eruption after magnetic reconnection around the hot cusp-structure above the top of plasma loops, however, the early evolution of the plasma loops before the onset of magnetic reconnection has been poorly understood. Considering that magnetic-gradients are ubiquitous in solar plasma loops, this work applies the magnetic-gradient pumping (MGP) mechanism to study the early evolution of flaring plasma loops. The results indicate that the early evolution depend on the magnetic field distribution and the geometry of the plasma loops which dominates the balance between the accumulation and dissipation of energy around loop-tops. Driven by MGP process, both of the density and temperature as well as the plasma $\beta$ value around the looptop will increase in the early phase of the plasma loop's evolution. In fact, the solar plasma loops will have two distinct evolutionary results: the low, initial dense plasma loops with relatively strong magnetic fields tend to be stable for their maximum $\beta$ value always smaller than the critical value $\beta < \beta_c$, while the higher, initial dilute solar plasma loops with relatively weak magnetic fields tend to be unstable for their $\beta$ values exceeding the critical value $\beta > \beta_c$ at a time of about one hour after the formation of the solar magnetized plasma loop. The latter may produce ballooning instability and finally trigger the following magnetic reconnection and eruptions. These physical scenarios may provide us a new viewpoint to understand the nature and origin of solar flares.

## 1. Introduction

Flares occurring on the Sun and other main-sequence stars are generated from the release of large amounts of energy and mass into the surrounding atmosphere and space. Solar flare is the most violent explosive process that occurs on the Sun – the nearest star to us in the universe. Although they have been studied for more than 160 years, many essential questions remain unresolved, such as what powers the eruptions? What is the primary trigger? Is there definite precursor to a flare? etc. The answers to these questions may help us to better predict when, where, and how solar flares occur, and avoid their damage to our Earth's space environment as soon as possible.

As we know, the solar flaring active region is always structured with various scales of plasma loops [1-3]. The standard flare models [4-8] proposed that magnetic reconnection around the cusp-shaped structure above the flaring plasma loops could effectively release magnetic energy, accelerate particles, heat and eject plasmas [9-11]. We may roughly divide the whole flaring process into three phases [12]: a long preflare phase, a fast-rising phase (also called impulsive phase), and a relatively long decay phase (postflare phase). We are nearly well understanding the evolution of the flaring plasma loops after the onset of magnetic reconnection [2, 11, 13-16]. Large number of observations and numerical simulations have demonstrated in detail the evolution of flaring plasma loops during their impulsive and postflare phases [17-19]. However, little is known about the early evolution of the flaring plasma loops during the preflare phase. Recently, Tan et al. (2020) applied the magnetic-gradient pumping (MGP) mechanism [20] to demonstrate the evolution of flaring plasma loops [21]: the magnetic-gradient force may drive energetic particle upflow which carries and conveys kinetic energy from the solar lower atmosphere with strong magnetic field to move upwards, accumulate and increase the temperature and plasma pressure around the looptop with relatively weak magnetic fields, produce plasma ballooning instability, and finally trigger magnetic reconnection and the following violent flaring eruption.

In fact, the whole solar corona is also highly structured where the building blocks are plasma loops. Some of them may become unstable and produce eruptions, while the others may be very stable for very long time on the Sun and do not erupt [2]. As we know that the magnetic-gradients are ubiquitous in nearly all coronal plasma loops, why do different loops have different evolutionary results? This work will attempt to answer this question. Section 2 presents analysis of the energy balance in coronal plasma loops. Section 3 discusses the evolution of coronal plasma loop during the early phase. Section 4 comes to conclusion of this work.

## 2. The analysis of energy balance of coronal plasma loops

2.1 Transport and accumulate areas of coronal plasma loops

Generally, a typical solar magnetized plasma loop is embedded on the photosphere, stretched downward into the solar hot interior with convection motions, and goes up into corona with height from one or two thousand to nearly million km [1-3]. In such plasma loops, footpoints near the photosphere are the strong magnetic field regions, while the looptops in the corona are the weak magnetic field regions. Therefore, there are magnetic gradients ($\nabla B$) in such plasma loops, and the direction of the gradient is always downward, see Figure 1. According to the MGP mechanism [20, 21`], the upward magnetic-gradient force ($F_m = -\mu \nabla B$, here $\mu = \frac{\frac{1}{2}mv_t^2}{B}$ is the magnetic moment) will drive energetic particles to move upward and form an energetic upflow in the plasma loop, extracts the energetic particles and kinetic energy from both footpoints of the loop, conveys and transports them to accumulate around the looptop. At the same time, the deficit of energetic particles near both footpoints can be replenished by the convection motion of the solar interior hot plasmas. The MGP is a continuous process, and the magnetized plasma loop is like a pumper which may continuously pump the energetic particles from the solar lower atmosphere to transport upward into the corona.

Naturally, the convection motion of the photosphere and chromosphere may produce various kinds of acoustic waves and MHD waves which possibly heat the upper part of the plasma loop, the interaction between the adjacent plasma loops may trigger magnetic reconnections and release magnetic energy. However, in this work we are not going to discuss the effects of acoustic waves and MHD waves and interactions between loops, but focus on discussing the energy transportation and conversion from the MGP mechanism. This is a completely different viewpoint and new approach to study the nature of solar flaring plasma loops and the related eruptions.

The downward magnetic-gradients near both footpoints are ubiquitous in all coronal plasma loops. But from the footpoint to the looptop, the magnetic-gradient is not uniform. Many observations can indirectly support this point [22-24]. Theoretically, from footpoint to leg of the loop the magnetic gradient is considerable, and the upward magnetic-gradient force should be stronger than the solar gravitational force which may pump the energetic particles to transport upward. But around looptop the magnetic gradient is insignificant and the upward magnetic-gradient force should be weaker than the solar gravitational force. In such case the energetic particles may accumulate around the looptop. Therefore, we may divide the whole solar magnetized plasma loop into two distinct areas: transport path and accumulate areas (shown in Figure 1).

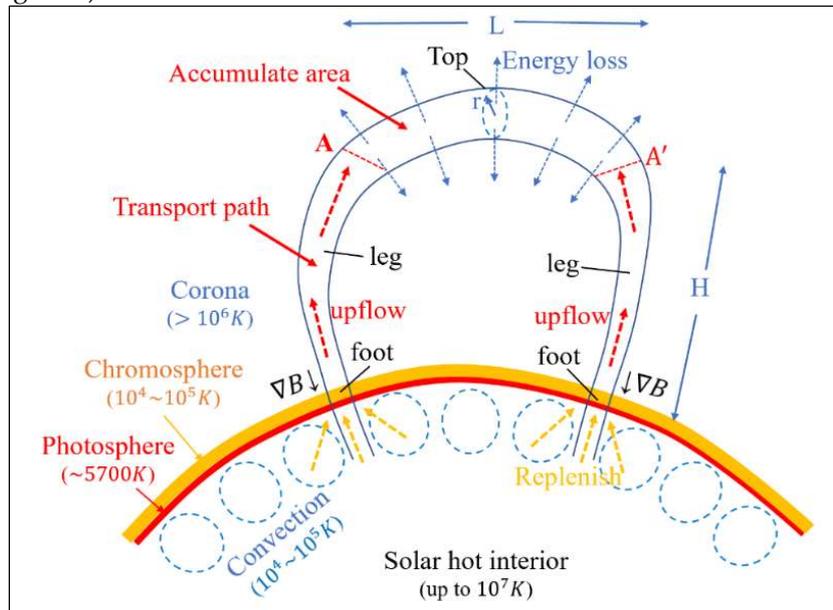

Figure 1. The schematic diagram of a solar magnetized plasma loop. Here the loop embeds on the photosphere and goes up to corona and down to the solar hot interior with convection motions. The directions of the magnetic-gradients point to downward at both footpoints which results in the energetic particle flows upward in the legs of the loop. The red dashed lines (A and A') are the roughly boundary between the transport path and the accumulate area.

(1) **Transport path**. The areas of the two legs of the loop below the red dashed lines (A and A' in Figure 1) till to the foot points have considerable downward magnetic gradient along the field line. The MGP process mainly works in this area. Here can be regarded as the transport path of the energetic upflows. The height of this area is marked as $H$ showing in Figure 1. In this region, the particles with energy higher than the starting energy ($\epsilon_{to} = mg(h)L_B cos\theta$) will be transported and move upward, while the particles with energy lower than $\epsilon_{to}$ will be confined and stay in the lower position of the loop. Generally, the solar gravitational acceleration $g(h)$ and the magnetic field scale length ($L_B = \frac{B}{\nabla B}$) are functions of the height ($h$) in the legs of the loop above the solar surface. Only the particles with energy around $\epsilon_{to}$ can stay at the height of ($h$) above the solar surface. Therefore, the temperature of the legs at height $h$ is completely dominated by the starting energy $\epsilon_{to}(h)$ which mainly depends on the relative magnetic gradient ($\frac{\nabla B}{B}$) and the solar gravitational acceleration $g(h)$.

$$T(h) \approx \frac{mg(h)}{e} L_B \qquad (1)$$

Here, $m$ is the mass of proton, and e is the electric charge of electron. The temperature $T(h)$ is in unit of eV which already contains the Boltzmann constant $k_B$. We approximately neglect the effect of the angle between the magnetic field line. The solar gravitational acceleration $g(h)$ decreases with respect to the height $h$ above the solar optical surface, $g(h) \approx \frac{274}{(1+h/R_s)^2}$, but the magnetic field scale length $L_B$ generally increases with respect to the height above the solar surface. $R_s$ is the solar optical radius ($6.963 \times 10^8 \, m$).

Generally, it is very difficult to obtain the real value of the magnetic field scale length $L_B$ directly from observations. We may simply adopt the magnetic field model of Dulk & McLean [25] to derive the approximated value of $L_B$, $B \approx 0.5(R_s/h)^{3/2}$, the unit of $B$ is Gauss, $L_B \approx \frac{2}{3}h$. Here, the valid range of this model is about $0.02R_s < h < 10R_s$. Then the temperature can be expressed as $T(h) \approx 1.90 \times 10^{-6} \frac{h}{(1+h/R_s)^2}$, (eV).

(2) **Accumulate area**. The area around the top of the loop above the red dashed lines (A and A') in Figure 1 has no considerable magnetic gradient along the field lines, and therefore the MGP process can be negligible. All the energetic particles coming from the lower underlying plasmas via transport path will deposit and bounce back and forth around the top of the loop. Therefore, we may refer to the region around looptop as the accumulation area of energetic particle upflows. Because the plasma temperature is simply a measure of the average kinetic energy of particles, with the continuously accumulating of energetic particles, the temperature of the plasma will exceed $T(H)$ and gradually increase in this area. Based on the magnetic field model of Dulk & McLean [25], we may obtain $L_B$ and then $T(H) > 1.90 \times 10^{-6} \frac{H}{(1+H/R_s)^2}$, (eV). Here, $H$ is the height of the looptop with unit of m. All the energetic particles with energy higher than $mg(H)L_B(H)$ will be droved to move upward and gathering in the accumulation area.

Naturally, if there is no energy loss around the accumulate area, the MGP process will make the plasma to become hotter and hotter, and the plasma beta will continuously increase, and eventually break through the confinement of magnetic fields and generate ballooning instability, magnetic reconnection and eruptions [21]. This is a natural mechanism which indicates that the released energy during solar flaring eruptions primarily comes from the solar interior, and the magnetized plasma loops just play a role of the transporting channel of energy and energetic particles.

However, besides the energy input in the looptop by MGP process, there also exists energy loss, including thermal radiation, heat conduction, and particle dissipation, etc. It is evident that the looptop will become unstable when the MGP energy input is higher than the energy loss, and the loop will be stable when the MGP energy input does not exceed the energy loss. Section 2.2 will discuss the energy input, and the energy loss will be diacussed in Section 2.3.

2.2 Energy input

The energy input refers to the energy carried by the energetic particle upflow crossing the plane A-A' and going into the accumulate area of the loop. Here, we do not consider the energy input driven by the MHD waves. We also neglect the contribution of neutrinos and neutrons in flaring plasma loops because they are just generated from the nuclear fusion in the core of the Sun, far from the solar flaring plasma loops occurring in the solar atmosphere.

Because the energetic particle upflow is driven by MGP process from the lower underlying plasmas and all their energy should exceed the starting energy at height of H, $\epsilon_t > mg(H)L_B(H)$. Therefore, the energy input is dominated by the relative magnetic gradient B/∇B at the boundary between the transport path and the accumulate area. The input energy per unit area per unit time flying across the above boundary is dominated by the number of the energetic particles and their upward velocities, it can be calculated by the following integration,

$$P_{in} = n(H) \int_{\epsilon_0}^{\infty} \epsilon_k \cdot \frac{1}{6} f(\epsilon_k) \cdot v \cdot d\epsilon_k, \quad (2)$$

Here, $\epsilon_k$ is the kinetic energy of the energetic particle at unit of eV, $n(H)$ is the particle density near the boundary between the transport path and the accumulate area. $f(\epsilon_k)$ is the energy distribution function of particles. Generally, we may suppose that the plasma is in thermodynamic equilibrium, $f(\epsilon_k) = \frac{\epsilon_k}{[T(H)]^2} \exp\left[-\frac{\epsilon_k}{T(H)}\right]$. $v$ is the vertical component of the velocity of energetic particles which can be approximated: $v \approx \left(\frac{2e\epsilon_k}{m}\right)^{1/2}$. Here, we assume that the distribution of energetic particles is isotropic, i.e. particles are equally likely to move in all six directions: up, down, left, right, front and back. Therefore, the probability of upward particles is $\frac{1}{6}$. The total energy input into the accumulate area per unit time from both transport paths (two legs of the loop) can be calculated from the following equation,

$$E_{in} = 2\pi r^2 P_{in} = 2\pi r^2 n(H) \left(\frac{2e}{m}\right)^{1/2} \int_{\epsilon_0}^{\infty} \frac{1}{6[T(H)]^2} \epsilon_k^{5/2} \exp\left[-\frac{\epsilon_k}{T(H)}\right] \cdot d\epsilon_k, \quad (3)$$

Here, $r$ is the radius of the loop section. The lower limit of the integral $\epsilon_0$ is the starting energy at the height (H) of the boundary between the transport path and the accumulate area, $\epsilon_0 = mg(H)L_B(H)$. $g(H) \approx \frac{274}{(1+H/R_s)^2}$. Then, we may obtain $\epsilon_0 \approx 2.85 \times 10^{-6} \frac{L_B(H)}{(1+H/R_s)^2}$.

The magnetic field scale length $L_B(H)$ depends on the gradient of magnetic fields in the loop. For example, we may simply adopt the model of Dulk & McLean [25] and obtain $L_B(H) \approx \frac{2}{3}H$, then $\epsilon_0 \approx 1.90 \times 10^{-6} \frac{H}{(1+H/R_s)^2}$ (eV). From Equation (1), we know that $T(H) = \epsilon_0$.

With the energy input into the accumulate area, both the temperature and density will gradually increase in the accumulate area. The temperature increase per unit time (also called heating rate) can be calculated,

$$T'_{heat} = \frac{2E_{in}}{3\pi r^2 L n(H)} \approx \frac{3.08 \times 10^3}{L \cdot \epsilon_0^2} \int_{\epsilon_0}^{\infty} \epsilon_k^{\frac{5}{2}} \exp\left(-\frac{\epsilon_k}{\epsilon_0}\right) d\epsilon_k \quad (4)$$

Here, the unit of heating rate $T'_{hea}$ is $eV \cdot s^{-1}$, the unit of $\epsilon_0$ and $\epsilon_k$ is eV, and L is at unit of m. Because the value of $\epsilon_0$ depends on the magnetic gradient at the boundary with height H between the transport path and accumulate area in the plasma loops, the heating rate is also dominated by the related magnetic gradient. At the same time, Equation (4) shows that the heating rate also depends on the geometric parameter (L) of the plasma loop.

It must be noted that the MGP upflow energetic particles which are transported into the accumulate area can be replenished continuously from the solar hot interior by convection motions, therefore, the energy input is a continuous process which may heat the accumulate area continuously, and the temperature will increase with the above energy input.

Accompanying with the temperature increase, due to the injection of the upflow energetic particles driven by MGP process into the accumulate area, the plasma density will also increase with time, which can be calculated by the following integration:

$$n_i = n(H)\{1 + t \times \frac{2.31 \times 10^3}{L \cdot \epsilon_0^2} \int_{\epsilon_0}^{\infty} \epsilon_k^{\frac{3}{2}} \exp\left(-\frac{\epsilon_k}{\epsilon_0}\right) d\epsilon_k\} \quad (5)$$

$n'_i = \frac{2.31 \times 10^3}{L \cdot \epsilon_0^2} \int_{\epsilon_0}^{\infty} \epsilon_k^{\frac{3}{2}} \exp\left(-\frac{\epsilon_k}{\epsilon_0}\right) d\epsilon_k$ is the injection rate of MGP upflow energetic particles. Here, $t$ is the time after the formation of the magnetized plasma loop in the solar atmosphere. $n(H)$ and $T(H)$ are the plasma density and temperature in the loop at the height of the boundary between the accumulate area and transport paths, respectively. The initial value of $n(H)$ can be obtained from models. The left panel of Figure 2 presents the temporal evolution of plasma density in two example cases at different heights, the results indicate that the plasma density linearly increases with time, and the increment may reach up to several times than the initial values $n(H)$ in the accumulate area driven by MGP process. Here, because of the confinement of magnetic fields, we neglect the particle escape due to diffusion before the onset of the eruption. Obviously, the accumulation of the MGP upflow energetic particles around the looptop will provide the material basis for the following flaring eruptions.

2.3 Energy loss

With the continuous MGP process, the plasma confined in the accumulate area around the looptop will become hotter and denser. Normally, the temperature around the looptop will exceed $10^6$ K, and the plasma density is about $10^{15} \sim 10^{16}$ $m^{-3}$ with magnetic field strength of about 10-100 Gauss [2]. Accompanying with the temperature increase, the energy loss will become more and more significant in the forms of radiation, heat conduction and particle dissipations, and these will result in cooling of the plasma around the looptop.

Because of the confinement of magnetic field and the low plasma density, the energy loss due to heat conduction and particle dissipation across the plasma loop can be negligible. The main contribution will come from radiation, including thermal bremsstrahlung emission from the collision among electrons-ions, electron cyclotron emission in magnetic field, excitation emission from the transition between different energy levels of ions or atoms, and the recombination radiations [26]. The more detailed simulations of the cooling of coronal plasma loops can be found from the serial researches of Bradshow and Cargill [27-29]. Here, we just present the simplified forms of the total contributions of each possible radiation process to derive a clear relationship between the radiative energy loss and the physical parameters (magnetic field strength, density, and temperature, etc.).

(1) **Cyclotron and gyro-synchrotron emission**. With the magnetic field in the plasma of the accumulate area, particle accelerations due to gyration around the magnetic field lines become dominant. Cyclotron emission in thermal plasma and gyro-synchrotron emission in very hot thermal plasma will provide the main contribution of radiative energy loss. Generally, it is very complicated to calculate the total energy of cyclotron and gyrosynchrotron emission in magnetized plasma loops. In a very dilute solar plasma with thermal equilibrium distribution, we may obtain the approximation of the total energy loss released in unit volume at all possible frequencies,

$$P_{cy} \approx 6.2 \times 10^{-20} n_e T_e B^2 (1 + 4.22 \times 10^{-13} T_e) \qquad (6)$$

$B$ is the magnetic field strength with unit at Tesla. $T_e$ is at eV. $P_{cy}$ is at $W \cdot m^{-3}$. Here, we find that the total energy of cyclotron and gyrosynchrotron emission increases with respect to the plasma density ($n_e$) and temperature ($T_e$), and rapidly increases with the magnetic field strength (B).

(2) **Bremsstrahlung emission**. In thermal plasma, the accelerations due to collisions between electrons and ions may result in bremsstrahlung emission. The total energy of bremsstrahlung emission at all possible frequencies in unit volume can be approximated,

$$P_{br} \approx 2.40 \times 10^{-38} n_e n_i Z^2 T_e^{1/2} \qquad (7)$$

Here, the unit of $P_{br}$ is at $W \cdot m^{-3}$, $n_e$ and $n_i$ are the density of electrons and ions respectively. $T_e$ is the plasma temperature with unit of eV. The contribution of thermal bremsstrahlung emission increases with respect to the plasma temperature and rapidly increases with the plasma density.

(3) **Excitation emission**. This is the radiation produced when electrons in excited state in atoms or ions jump down to some lower energy state (transition of the energy levels), which may form a series of line emission. In solar coronal plasmas, it may contribute the energy loss in unit volume per unit time,

$$P_e \approx 2.7 \times 10^{-2} Z^4 T_e^{-2} P_{br} \qquad (8)$$

Here, the unit of $T_e$ is eV. $Z$ is the charge number of nuclei. Equation (8) indicates that the excitation emission decreases rapidly with increment of plasma temperature. At the same time, it is fast increase with the charge number Z. Typically in the plasma around the solar coronal looptop, $Z \sim 1$, $T_e \approx 100 - 1000$ eV, therefore $P_e \ll P_{br}$.

(4) **Recombination radiation**. When a free electron collides with ions, recombination may take place and radiate photons. The energy loss of recombination radiation in unit volume solar plasma per unit time can be approximated,

$$P_{rc} \approx \frac{R_y}{T_e} \exp\left(\frac{R_y}{T_e}\right) \cdot P_{br} \qquad (9)$$

Here, $R_y$ is the ionization energy of hydrogen. For ground state of hydrogen, $R_y \approx 13.58$ eV, and the typical plasma temperature (>100 eV) in the looptop $T_e \gg R_y$. Therefore, normally $P_{rc} \ll P_{br}$. Equation (9) indicate that the contributions of recombination radiation also decrease rapidly with the increase of plasma temperature.

The total energy loss due to radiations is the sum of Equations (6) – (9). However, because $P_e \ll P_{br}$ and $P_{rc} \ll P_{br}$, the emission contributions due to excitation and recombination are much smaller than the cyclotron and bremsstrahlung emission in typical coronal looptop plasma, the total energy loss in unit volume per unit time can be approximated into,

$$P_{out} \approx 6.2 \times 10^{-20} n_e T_e B^2 + 2.40 \times 10^{-3} n_e n_i T_e^{\frac{1}{2}} \qquad (10)$$

The total energy loss in the accumulation area can be obtained from the product of $E_{out} = P_{out} \cdot V$. Here, $V$ is the volume of the accumulate area of the plasma loop, $V \approx \pi r^2 L$. $L$ is the length of the accumulate area (showed in Figure 1). With the total energy loss, the plasma will be cooling and the temperature will decrease. The temperature decreases per unit time (cooling rate) can be estimated,

$$T'_{cool} = \frac{E_{out}}{\pi r^2 Ln(H)e} \approx 0.38 T_e B^2 + 1.5 \times 10^{-19} n_i T_e^{\frac{1}{2}} \qquad (11)$$

Here, we set the unit of $T_e$ as eV. It is very convenient for our following calculations. The unit of cooling rate $T'_{cool}$ is at $eV \cdot s^{-1}$, $B$ is at Tesla, and $n_i$ is at $m^{-3}$, $Z \approx 1$. $n_i$ can be obtained from Equation (5).

2.4 The temporal evolution of temperature

Equation (11) indicates that the radiative cooling of the accumulate area depends largely on the temperature ($T_e$), magnetic field strength ($B$), and the density ($n_i$) of the plasma. With Equation (4) and (11), we may obtain the total temporal evolution of the plasma temperature in the accumulate area,

$$T_e = T(H) + t \times (T'_{heat} - T'_{cool}), \qquad (12)$$

The above equation is an implicit expression of the temperature $T_e$, which can be solved,

$$T_e = X^2, \qquad (13)$$

$X = [-P + \sqrt{P^2 + 4Q(\epsilon_0 + T'_{heat}t)}]/(2Q)$, $P = 1.5 \times 10^{-1} n_i t$, $Q = 1 + 0.38 B^2 t$. $n_i$ can be obtained from Equation (5). Here, we assume two example loops with height of $H = 2 \times 10^7$ m and initial plasma density $n(H) = 1 \times 10^{16} m^{-3}$ (case1: black curve in Figure 2) and $H = 3 \times 10^7$ m and $n(H) = 0.8 \times 10^{16} m^{-3}$ (case2: red curve in Figure 2), respectively. Such assumption is basically consistent with the classical models of solar atmosphere [30, 31]. The middle panel of Figure 2 presents the temporal evolution of the temperature in the accumulation areas which shows that in case1 the temperature can gradually increase to $1.76 \times 10^6$ K (black curve and $T_1$ in the middle panel of Figure 2) at about 46 min after the formation of the magnetized plasma loop and then decrease slowly because of the radiation cooling. In case2, the temperature gradually increases to a maximum of $3.00 \times 10^6$ K (red curve and $T_2$ in the middle panel of Figure 2) at about 66 min after the formation of the loop, and then also slowly decreases. These maximum temperatures are consistent with our common observations of the coronal plasma loops [1-4].

**3. The evolution of coronal plasma loop during the early phase**

The energy input driven by MGP process from the lower underlying plasmas (Equation 3) will make both temperature and density in the accumulate area around the looptop to increase gradually, and the heating rate can be calculated from Equation (4). At the same time, with the increases of temperature and density in the plasma loop, the radiative loss will also gradually increase, causing the above plasma loop to cool. Here, there is a balance between the energy input and loss. If the energy input exceeds the energy loss, the plasma loop will be heated continuously till to trigger some instabilities. However, if the energy loss exceeds the energy input, the plasma will be cooling and the loop will not evolve to a stage of instability.

Here, we do not consider the interaction between plasma loops, and shear or twist motions which will certainly cause the loop's instabilities and eruptions [32-35]. We mainly investigate the static plasma loops with magnetic gradient. The important parameter to describe the stability of magnetized plasma loop is the $\beta$ value [36, 37], which is defined as the ratio between plasma thermal pressure and magnetic pressure. The thermal pressure reflects the expansion tendency of the plasma, while the magnetic pressure reflects the constraint ability of the magnetic field to the plasma. Therefore, for a given magnetized plasma loop, high $\beta$ value means that the plasma loop tends to get rid of the confinement of the magnetic field, while the low $\beta$ value implies that the plasma loop tends to be stable. $\beta$ value can be calculated by the following expression,

$$\beta = \frac{2\mu_0 e}{B^2} n_i T_e \approx 4.04 \times 10^{-25} \frac{n_i T_e}{B^2}, \qquad (14)$$

Here, the unit of $T_e$ is eV. Using Equation (5) we may obtain the plasma density $n_i$, and Equation (12) may give the plasma temperature in the accumulation area of the plasma loop. We may approximately adopt the model of Dulk & McLean [25] to derive the magnetic field ($B$) and its scale length ($L_B$), and the starting energy ($\epsilon_{to}$) of the upward energetic particles at certain height H.

For a given magnetized plasma loop, there exists a critical $\beta$ value ($\beta_c$) [38, 39]. When $\beta \geq \beta_c$, the magnetic field cannot confine the plasma and the loop will evolve to produce instability. Tsap et al. found [40, 41] that the critical value $\beta_c$ for ballooning instability mainly depends on geometrical parameters of the plasma loop, $\beta_c \approx \frac{2r}{R} \approx \frac{2r}{H}$. Here, R is the radius of the loop's curvature and r is the section radius of the loop. In the case of Figure 1, $R \sim H$. Numerous observations indicate that the ratio $\frac{2r}{R}$ is about 0.05-0.1 in most coronal loops [2, 3], therefore, $\beta_c \approx 0.1$.

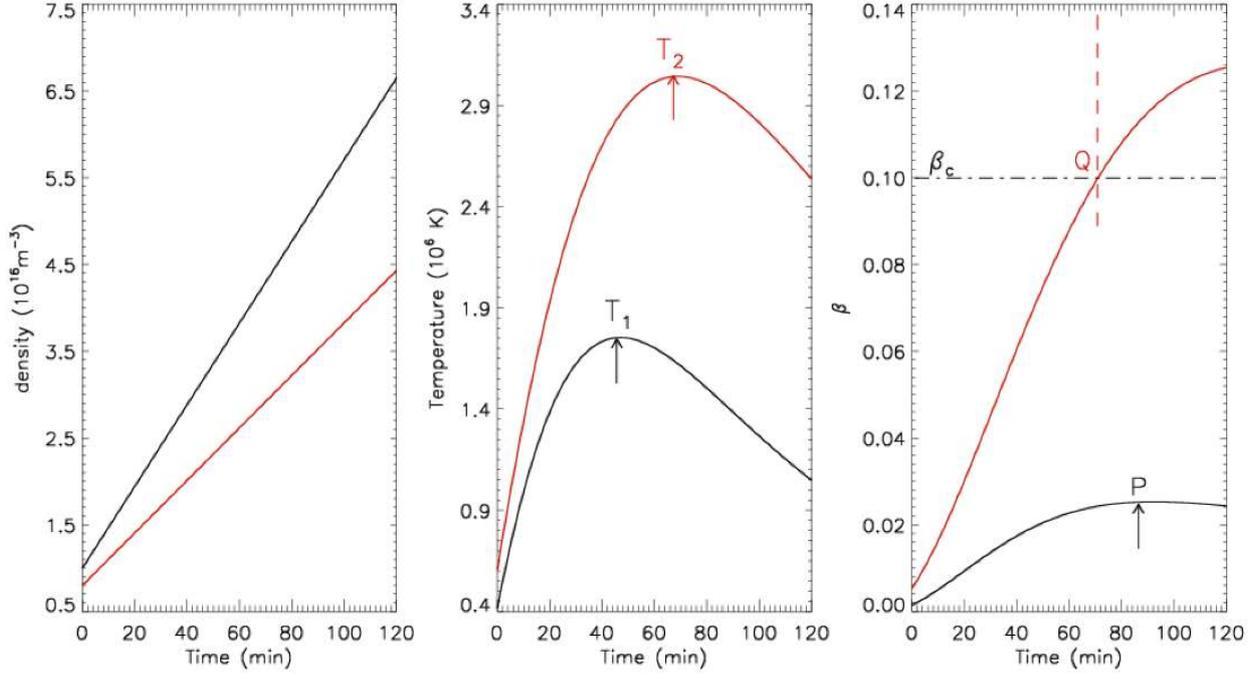

**Figure 2.** The temporal evolution of the plasma density (left), temperature (middle) and $\beta$ value (right) in two cases. In case 1 (black curves), we assume the initial density $n(H) = 1.0 \times 10^{16} \ m^{-3}$ at height H= $2 \times 10^7$m. In case 2 (red curves), we assume the initial density $n(H) = 0.8 \times 10^{16} \ m^{-3}$ at height H= $3 \times 10^7$m. Here, we adopt the model of magnetic field of Dulk & McLean [25] to derive the magnetic field (about 101 Gauss in case 1 and 56 Gauss in case 2), magnetic scale length ($L_B$) and the starting energy ($\epsilon_o$, 35.9 eV in case 1 and 52.3 eV in case 2) of the upward energetic particles.

We calculate the temporal evolutions of $\beta$ value in the accumulate area of coronal plasma loops at certain heights (H). In practice, we don't know the true plasma densities $n(H)$ near the boundary between the transport path and the accumulate area. Here, we simply assume the values of $n(H)$ empirically as $1.0 \times 10^{16} \ m^{-3}$ at height of $2.0 \times 10^7$m (case 1) and $0.8 \times 10^{16} \ m^{-3}$ at height of $3.0 \times 10^7$m (case 2) as examples. This assumption is consistent with the typical models and observations of solar atmosphere [30, 31] Based on the mentioned magnetic field model, we may figure out the corresponding magnetic field strength as 101 Gauss in case 1 and 56 Gauss in case 2. The results are presented in the right panel of Figure 2.

In case 1, the $\beta$ value (the black curve) increases with time and reaches to a maximum of about 0.025 at about 86 min (P point in Figure 2) after the formation of the magnetized plasma loop, then it decreases slowly. Here, we note that the maximum $\beta$ is only 0.025, which is much smaller than the possible critical value ($\beta_c \approx 0.10$). That is to say, the plasma loop will remain stable at all time.

However, in case 2, the $\beta$ value (the red curve) exceeds the possible critical value at about 70 min ($\beta > \beta_c$, Q point in Figure 2) after the formation of the magnetized plasma loop. That is to say, the plasma loop may trigger a ballooning instability after Q point. Here, it is necessary to note that the above calculation of $\beta$ is an averaged result in the accumulation area. However, in fact, this area is not homogeneous. For example, the magnetic field strength near the looptop may be weakest and therefore here the $\beta$ value may be the biggest. Naturally, the ballooning instability and the following eruptions may take place from the looptop at first [42].

The comparison between case 1 and case 2 implies that lower, initial dense solar plasma loop with relatively strong magnetic fields tends to be stable, while the higher, dilute solar plasma loop with relatively weak magnetic field tend to be unstable and may generate eruptions.

In order to further illustrate the effect of the initial density on the temporal evolution of the plasma loops, we assume that the accumulate area has different initial density ($0.8 \times 10^{16} \ m^{-3}$, $1.2 \times 10^{16} \ m^{-3}$, and $2.0 \times 10^{16} \ m^{-3}$) at the same height ($3 \times 10^7$m), and then calculate the evolutions of the density, temperature and $\beta$ value. The results are presented in Figure 3. Here, we also found that the dense magnetized plasma loop tends to become more cold, more dense and stable (the red curves in Figure 3), while the dilute magnetized plasma loop tends to be unstable (the black curves in Figure 3).

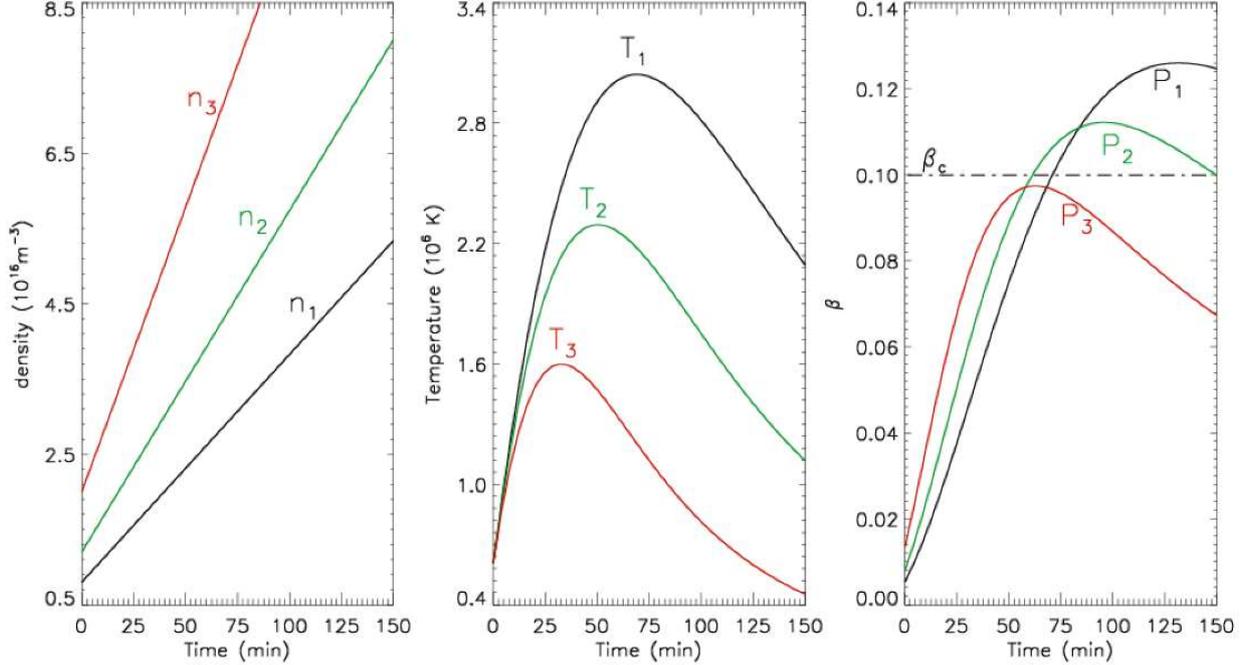

**Figure 3.** The temporal evolution of the plasma density (left), temperature (middle) and $\beta$ value (right) with different initial density at the same height H = $3 \times 10^7$ m. The initial density $n(H) = 0.8 \times 10^{16}\ m^{-3}$ (black), $1.2 \times 10^{16}\ m^{-3}$ (blue) and $2.0 \times 10^{16}\ m^{-3}$ (red), respectively. We also adopt the model of magnetic field of Dulk & McLean [25] to derive the magnetic field (about 56 Gauss), magnetic scale length ($L_B$) and the starting energy (52.3 eV) of the upward energetic particles.

In the above calculations, the initial plasma density $n(H)$ is just assumed empirically, and the magnetic field is simply derived from the model of Dulk & McLean [25]. In practice, we can also calculate and present the temporal evolutions of solar coronal loops by adopting the more accurate models (generally the more accurate model is just the more complicated one) and even the observational data to derive the initial parameters (including density, magnetic field, scale length and the starting energy).

## 4. Conclusions

From the above analysis by using the MGP mechanism, we may draw the following conclusions for the early evolution of solar flaring plasma loops.

(1) The solar plasma loop can be divided into two distinct areas: transport path (from foot-points to the loop legs with considerable magnetic gradient) and accumulate area (around the looptop). The energetic particles comprising in the underlying thermal plasma can be driven by MGP process to flow upward through transport paths and deposit in the accumulate area.

(2) With the deposit and accumulation of MGP upflow energetic particles, the plasma density in accumulate area may increase several times than the initial value in about one or two hours. The accumulation of the MGP upflow energetic particles around the looptop provides the material basis for the possible following flaring eruptions.

(3) The plasma temperature in the accumulate area gradually increases to a maximum of several million K in several decades minutes after the formation of the loops. Then it decreases slowly due to the enhancement of the bremsstrahlung and cyclotron radiations.

(4) The $\beta$ value is a key parameter to show the temporal evolutionary features of a plasma loop. We find that the $\beta$ value of the accumulate area also gradually increases driven by MGP process in the first decades of minutes. However, in fact, not all solar coronal loops can evolve into an unstable stage to produce eruptions. Most of the low, initially dense solar plasma loops with relatively strong magnetic fields tend to be stable (such as case 1 showing in Figure 2) for the maximum $\beta$ value still much smaller than the critical value $\beta_c$. This can be demonstrated the evolution of many quiet, or quasi-static plasma loops in the solar atmosphere [32]. On the other hand, the higher, initially dilute solar plasma loop with relatively weak magnetic fields tends to be unstable (such as case 2 showing in Figure 2) for the $\beta$ value exceeding the critical value $\beta > \beta_c$ at a time of about one hour after the formation of the solar magnetized plasma loop which may trigger ballooning instability and the following eruptions. This case can be applied to demonstrate the early evolution of flaring plasma loops [43, 44].

As for the flaring plasma loop, because it will eventually evolve into eruptions, its early evolutionary stage can be driven by MGP process to increase its density, temperature, and the $\beta$ value to exceed the critical value and trigger a ballooning instability which may generate suitable conditions for the following magnetic reconnections [21]. And such early stage may last for several decades of minutes. This result may answer the question raised at the beginning of this article: the flaring plasma loop may go through a precursor phase which lasts for about one hour or so before the onset of the following eruptions. During this phase, the plasma density and temperature in the area around the looptop will gradually increase. And the corresponding observational characteristics are mainly reflected by the enhancement of bremsstrahlung and cyclotron radiations, and possibly produce some unique observed phenomena, such as the very long period QPP [11, 45, 46] or other precursors.

The merits of MGP mechanism for understanding the early evolution of flaring plasma loops is that it provides a self-consistent link between the solar interior motion, transport path and the flaring source region. At the same time, the MGP process extracts the energetic particles from the solar underlying atmosphere which is replenished from the convection motion of the solar interior plasmas, then the energetic particles deposit and accumulate around the looptop. This physical regime may provide main conditions of mass and energy for the subsequent eruptions. We will continue to investigate the MGP mechanism in the solar and stellar coronal loops for atmospheric heating and eruptions in the following works by using the multi-wavelength observations, including the radio broadband spectral-imaging, such as MUSER [47], and the multi-wavelength EUV imaging, such as AIA/SDO [48], etc.

**Funding:** This research was funded by the National Natural Science Foundation of China, grant number 11790301, 11973057, 11941003, and the MOST Key Project 2018YFA0404602.

**Acknowledgments:** The author would like to thank the referees for their helpful and valuable comments to improve the manuscript of this paper. This work is also supported by the international collaboration of ISSI-BJ and the International Partnership Program of Chinese Academy of Sciences (grant number 183311KYSB20200003). The author thanks Prof. Guangli Huang and Dr. Yin Zhang for their valuable discussions.

**Conflicts of Interest:** The authors declare no conflict of interest. The funders had no role in the design of the study; in the collection, analyses, or interpretation of data; in the writing of the manuscript, or in the decision to publish the results.